\documentclass[apjl,iop,twocolappendix,numberedappendix]{emulateapj}

\usepackage{relsize}
\usepackage{color}
\usepackage{xcolor}
\usepackage{amsmath}
\usepackage{mathtools}
\usepackage{multirow}
\usepackage{hyperref}
\usepackage{graphicx}

\newcommand{\ddt}[1]{\frac{\partial{#1}}{\partial t}}
\newcommand{\mach}{\mathcal{M}}
\newcommand{\cs}{c_\mathrm{s}}
\newcommand{\bfu}{\mathbf{v}}
\newcommand{\bfB}{\mathbf{B}}
\newcommand{\re}{\mathrm{Re}}
\newcommand{\rmag}{\mathrm{Rm}}
\newcommand{\pmag}{\mathrm{Pm}}
\newcommand{\emag}{E_\mathrm{mag}}
\newcommand{\ekin}{E_\mathrm{kin}}
\newcommand{\esat}{(\emag/\ekin)_\mathrm{sat}}

\slugcomment{The Astrophysical Journal Letters, in press, \today}

\shorttitle{The Turbulent Dynamo}
\shortauthors{Federrath et al.}

\begin{document}

\title{The Turbulent Dynamo in Highly Compressible Supersonic Plasmas}

\author{Christoph~Federrath\altaffilmark{1}, Jennifer~Schober\altaffilmark{2}, Stefano~Bovino\altaffilmark{3}, Dominik~R.~G.~Schleicher\altaffilmark{3}}
\email{christoph.federrath@anu.edu.au}

\altaffiltext{1}{Research School of Astronomy and Astrophysics, The Australian National University, Canberra, ACT~2611, Australia}
\altaffiltext{2}{Universit\"at Heidelberg, Zentrum f\"ur Astronomie, Institut f\"ur Theoretische Astrophysik, Albert-Ueberle-Strasse~2, D-69120 Heidelberg, Germany}
\altaffiltext{3}{Institut f\"ur Astrophysik, Georg-August-Universit\"at G\"ottingen, Friedrich-Hund-Platz~1, D-37077 G\"ottingen, Germany}

\begin{abstract}
The \emph{turbulent dynamo} may explain the origin of cosmic magnetism. While the exponential amplification of magnetic fields has been studied for incompressible gases, little is known about dynamo action in highly compressible, supersonic plasmas, such as the interstellar medium of galaxies and the early Universe. Here we perform the first quantitative comparison of theoretical models of the dynamo growth rate and saturation level with three-dimensional magnetohydrodynamical simulations of supersonic turbulence with grid resolutions of up to $1024^3$ cells. We obtain numerical convergence and find that dynamo action occurs for both low and high magnetic Prandtl numbers \mbox{$\pmag=\nu/\eta=0.1$--$10$} (the ratio of viscous to magnetic dissipation), which had so far only been seen for $\pmag\geq1$ in supersonic turbulence. We measure the critical magnetic Reynolds number, $\rmag_\mathrm{crit}=129^{+43}_{-31}$, showing that the compressible dynamo is almost as efficient as in incompressible gas. Considering the physical conditions of the present and early Universe, we conclude that magnetic fields need to be taken into account during structure formation from the early to the present cosmic ages, because they suppress gas fragmentation and drive powerful jets and outflows, both greatly affecting the initial mass function of stars.
\end{abstract}


\section{Introduction}

Magnetic field amplification via the \emph{turbulent dynamo} is believed to be the main cause of cosmic magnetism. The turbulent dynamo is important for the formation of the large-scale structure of the Universe \citep{RyuEtAl2008}, in clusters of galaxies \citep{SubramanianShukurovHaugen2006} and in the formation of the first cosmological objects \citep{SchleicherEtAl2010}. It determines the growth of magnetic energy during solar convection \citep{CattaneoHughes2001,MollEtAl2001,GrahamEtAl2010}, in the interior of planets \citep{RobertsGlatzmaier2000} and in liquid metal experiments on Earth \citep{MonchauxEtAl2007}. It may further explain the far-infrared-radio correlation in spiral galaxies \citep{SchleicherBeck2013}. After the turbulent dynamo has amplified tiny seeds of the magnetic field, which can be generated during inflation, the electroweak or the QCD phase transition \citep{GrassoRubinstein2001}, the large-scale dynamo kicks in and generates the large-scale magnetic fields that we observe in planets, stars and galaxies today \citep{BeckEtAl1996,BrandenburgSubramanian2005}.

The properties of the turbulent dynamo strongly depend on the magnetic Prandtl number, $\pmag=\nu/\eta$, defined as the ratio of viscosity $\nu$ to magnetic diffusivity $\eta$ \citep{SchekochihinEtAl2004}. On large cosmological scales and in the interstellar medium, we typically have $\pmag\gg1$, while for the interior of stars and planets, the case with $\pmag\ll1$ is more relevant \citep{SchekochihinEtAl2007}. Numerical simulations, on the other hand, are typically restricted to $\pmag\sim1$, because of limited numerical resolution. Simulations by \citet{IskakovEtAl2007} have clearly demonstrated that the turbulent dynamo operates for $\pmag\lesssim1$ in incompressible gases, even though an asymptotic scaling relation has not been confirmed. While the bulk of previous work was dedicated to exploring the turbulent dynamo in the incompressible regime \citep{BrandenburgSokoloffSubramanian2012}, most astrophysical systems show signs of high compressibility. This is particularly true during the formation of the first cosmological objects \citep{LatifSchleicherSchmidt2014}, in the interstellar medium of galaxies \citep{Larson1981} and in the intergalactic medium \citep{IapichinoVielBorgani2013}. The compressibility of the plasma can be characterized in terms of the sonic Mach number $\mach=V/\cs$, the ratio of the turbulent velocity $V$ and the sound speed $\cs$. The Mach number typically exceeds unity by far in all of these systems, which is a hallmark of highly compressible, supersonic turbulence.

In the framework of the Kazantsev model \citep{Kazantsev1968}, \citet{SchoberEtAl2012PRE2} derived analytical dynamo solutions for the limiting cases $\pmag\to \infty$ and $\pmag\to0$, considering different scaling relations of the turbulence, while \citet{BovinoEtAl2013} derived a numerical solution of the Kazantsev equation for finite values of $\pmag$. These studies strongly suggest that the turbulent dynamo operates for different values of $\pmag$, as long as the magnetic Reynolds number, $\rmag=VL/\eta$, is sufficiently high, where $L$ is the characteristic size of the large-scale turbulent structures.

However, a central restriction of the Kazantsev framework is the assumption of an incompressible velocity field, for which a separation into solenoidal and compressible parts is not necessary. The distinction between solenoidal and compressible modes, however, may be essential for highly compressible, supersonic turbulence. Furthermore, the Kazantsev framework assumes that the turbulence is $\delta$-correlated in time, which is not appropriate for real turbulence. The resulting uncertainties introduced by the last assumption, however, are only a few percent \citep{SchekochihinKulsrud2001,KleeorinRogachevskiiSokoloff2002,BhatSubramanian2014}, while the assumption of incompressibility is a severe limitation. Ultimately, the full non-linear solution through three-dimensional (3D) simulations is needed to determine the behavior of the growth rates under more realistic conditions.

We note that the turbulent dynamo has also been studied in the context of so-called shell models \citep[][and references therein]{FrickEtAl2006}. These approaches allow us to derive theoretical predictions for the magnetic field growth in the exponential and in the saturated regime, and are highly complementary to the methods presented here.

In this Letter, we present the first investigation of the turbulent dynamo and its dependence on the magnetic Prandtl number in the highly compressible, supersonic regime. For this purpose, we consider supersonic turbulence with Mach numbers ranging from $\mach=3.9$ to $11$, and magnetic Prandtl numbers between $\pmag=0.1$ and $10$. The results are compared with the predictions from the Kazantsev model. Section~\ref{sec:sims} defines the numerical methods used in the simulations, Section~\ref{sec:theory} summarizes current dynamo theories, Sections~\ref{sec:results} and~\ref{sec:conclusions} present our results and conclusions.

\section{Numerical simulations} \label{sec:sims}

We use a modified version of the FLASH code \citep{FryxellEtAl2000} (v4) to integrate the 3D, compressible, magnetohydrodynamical (MHD) equations, including viscous and resistive dissipation terms,
\begin{align}
& \ddt\,\rho + \nabla\cdot\left(\rho \bfu\right)=0, \label{eq:mhd1} \\
& \! \begin{multlined}
	\ddt\!\left(\rho \bfu\right) + \nabla\cdot\left(\rho \bfu\!\otimes\!\bfu - \frac{1}{4\pi}\bfB\!\otimes\!\bfB\right) + \nabla p_\mathrm{tot} = {} \\
	\nabla\cdot\left(2\nu\rho\boldsymbol{\mathcal{S}}\right) + \rho{\bf F},
\end{multlined} \label{eq:mhd2} \\
& \! \begin{multlined}
	\ddt\,E + \nabla\cdot\left[\left(E+p_\mathrm{tot}\right)\bfu - \frac{1}{4\pi}\left(\bfB\cdot\bfu\right)\bfB\right] = {} \\
	\nabla\cdot\left[2\nu\rho\bfu\cdot\boldsymbol{\mathcal{S}}+\frac{1}{4\pi}\bfB\times\left(\eta\nabla\times\bfB\right)\right],
\end{multlined} \label{eq:mhd3} \\
& \ddt\,\bfB = \nabla\times\left(\bfu\times\bfB\right) + \eta\nabla^2\bfB, \label{eq:mhd4} \\
& \nabla\cdot\bfB = 0. \label{eq:mhd5}
\end{align}
In these equations, $\rho$, $\bfu$, $p_\mathrm{tot}=p_\mathrm{th}+ (1/8\pi)\left|\bfB\right|^2$, $\bfB$, and $E=\rho \epsilon_\mathrm{int} + (1/2)\rho\left|\bfu\right|^2 + (1/8\pi)\left|\bfB\right|^2$ denote plasma density, velocity, pressure (thermal plus magnetic), magnetic field, and energy density (internal plus kinetic, plus magnetic), respectively. Physical shear viscosity is included via the traceless rate of strain tensor, $\mathcal{S}_{ij}=(1/2)(\partial_i u_j+\partial_j u_i)-(1/3)\delta_{ij}\nabla\cdot\bfu$ in the momentum Equation~(\ref{eq:mhd2}), and controlled by the kinematic viscosity, $\nu$. Physical diffusion of $\bfB$ is controlled by the magnetic resistivity $\eta$ in the induction Equation~(\ref{eq:mhd4}). To solve the MHD equations, we use the positive-definite second-order accurate HLL3R Riemann scheme, capable of handling strong shocks \citep{WaaganFederrathKlingenberg2011}. The MHD equations are closed with an isothermal equation of state, $p_\mathrm{th}=\cs^2\rho$.

To drive turbulence with a given Mach number $\mach$, we apply a divergence-free large-scale forcing term ${\bf F}$ as a source term in the momentum Equation~(\ref{eq:mhd2}). The forcing is modeled with a stochastic Ornstein-Uhlenbeck process \citep{FederrathDuvalKlessenSchmidtMacLow2010}, such that ${\bf F}$ varies smoothly in space and time with an auto-correlation equal to the eddy-turnover time, $T=L/(2\mach\cs)$ on the largest scales, $L/2$ in our periodic simulation domain of side length $L$.

The efficiency of magnetic field amplification depends on the growth rate, which in turn depends on the driving mode, the Mach number, the Reynolds numbers $\re$ and $\rmag$, and the Prandtl number, $\pmag$ \citep{FederrathEtAl2011PRL,SchoberEtAl2012PRE2,BovinoEtAl2013,SchleicherEtAl2013}. We run most of our simulations until saturation of the magnetic field is reached. Given the Reynolds numbers achievable in state-of-the-art simulations, this can take several hundred crossing times. Saturation occurs when the Lorentz force induces a back reaction of the magnetic field strong enough to counteract the turbulent twisting, stretching and folding of the field \citep{BrandenburgSubramanian2005}. We determine the saturation levels by measuring the ratio of magnetic to kinetic energy, $\esat$.

Here we study the dependence of the turbulent dynamo on $\pmag$, which is accomplished by varying the physical viscosity and resistivity. Table~\ref{tab:sims} provides a complete list of all simulations and key parameters. To test convergence, we run simulations with \mbox{$N_\mathrm{res}^3=128^3$--$1024^3$} grid points.

\begin{table*}
\caption{List of Turbulent Dynamo Simulations}
\label{tab:sims}
\def\arraystretch{0.5}
\begin{tabular*}{\linewidth}{@{\extracolsep{\fill} }lccccccc}
\hline
\hline
Simulation Model & $N_\mathrm{res}^3$ & $\mach$ & $\pmag$ & $\re$ & $\rmag$ & $\Gamma\;(T^{-1})$ & $\esat$ \\
(1) & (2) & (3) & (4) & (5) & (6) & (7) & (8) \\
\hline
01) \texttt{Dynamo\_512\_Pm0.1\_Re1600} & $ 512^3$ & $ 11$ & $0.1$ & $ 1600$ & $  160$ & $(  2.7\pm3.0)\!\times\!10^{-3 }$ & n/a \\
02) \texttt{Dynamo\_1024\_Pm0.1\_Re1600} & $1024^3$ & $ 11$ & $0.1$ & $ 1600$ & $  160$ & $(  1.9\pm 50)\!\times\!10^{-3 }$ & n/a \\
03) \texttt{Dynamo\_512\_Pm0.2\_Re1600} & $ 512^3$ & $ 11$ & $0.2$ & $ 1600$ & $  320$ & $(  3.5\pm0.4)\!\times\!10^{-2 }$ & $(  6.0\pm2.0)\!\times\!10^{-4 }$ \\
04) \texttt{Dynamo\_512\_Pm0.5\_Re1600} & $ 512^3$ & $ 11$ & $0.5$ & $ 1600$ & $  810$ & $(  2.0\pm0.2)\!\times\!10^{-1 }$ & $(  1.0\pm0.3)\!\times\!10^{-2 }$ \\
05) \texttt{Dynamo\_512\_Pm2\_Re1600} & $ 512^3$ & $ 11$ & $  2$ & $ 1600$ & $ 3200$ & $(  4.5\pm0.4)\!\times\!10^{-1 }$ & $(  3.0\pm1.0)\!\times\!10^{-2 }$ \\
06) \texttt{Dynamo\_256\_Pm5\_Re1600} & $ 256^3$ & $ 11$ & $  5$ & $ 1600$ & $ 8000$ & $(  6.4\pm0.6)\!\times\!10^{-1 }$ & $(  3.9\pm1.3)\!\times\!10^{-2 }$ \\
07) \texttt{Dynamo\_512\_Pm5\_Re1600} & $ 512^3$ & $ 11$ & $  5$ & $ 1600$ & $ 8000$ & $(  5.8\pm0.6)\!\times\!10^{-1 }$ & $(  5.2\pm1.7)\!\times\!10^{-2 }$ \\
08) \texttt{Dynamo\_1024\_Pm5\_Re1600} & $1024^3$ & $ 11$ & $  5$ & $ 1600$ & $ 8000$ & $(  6.2\pm0.6)\!\times\!10^{-1 }$ & $(  4.6\pm1.5)\!\times\!10^{-2 }$ \\
09) \texttt{Dynamo\_256\_Pm10\_Re1600} & $ 256^3$ & $ 11$ & $ 10$ & $ 1600$ & $16000$ & $(  6.9\pm0.7)\!\times\!10^{-1 }$ & $(  4.0\pm1.3)\!\times\!10^{-2 }$ \\
10) \texttt{Dynamo\_512\_Pm10\_Re1600} & $ 512^3$ & $ 11$ & $ 10$ & $ 1600$ & $16000$ & $(  6.4\pm0.6)\!\times\!10^{-1 }$ & $(  5.7\pm1.9)\!\times\!10^{-2 }$ \\
11) \texttt{Dynamo\_1024\_Pm10\_Re1600} & $1024^3$ & $ 11$ & $ 10$ & $ 1600$ & $16000$ & $(  6.5\pm0.6)\!\times\!10^{-1 }$ & $(  4.8\pm1.6)\!\times\!10^{-2 }$ \\
12) \texttt{Dynamo\_128\_Pm10\_Re4.7} & $ 128^3$ & $4.0$ & $ 10$ & $  4.7$ & $   47$ & $(  6.0\pm160)\!\times\!10^{-3 }$ & n/a \\
13) \texttt{Dynamo\_256\_Pm10\_Re4.6} & $ 256^3$ & $3.9$ & $ 10$ & $  4.6$ & $   46$ & $(  5.8\pm160)\!\times\!10^{-3 }$ & n/a \\
14) \texttt{Dynamo\_128\_Pm10\_Re15} & $ 128^3$ & $6.4$ & $ 10$ & $   15$ & $  150$ & $(  3.4\pm0.6)\!\times\!10^{-2 }$ & n/a \\
15) \texttt{Dynamo\_256\_Pm10\_Re15} & $ 256^3$ & $6.4$ & $ 10$ & $   15$ & $  150$ & $(  4.3\pm0.7)\!\times\!10^{-2 }$ & n/a \\
16) \texttt{Dynamo\_128\_Pm10\_Re26} & $ 128^3$ & $7.5$ & $ 10$ & $   26$ & $  260$ & $(  2.9\pm0.3)\!\times\!10^{-1 }$ & $(  4.1\pm1.6)\!\times\!10^{-2 }$ \\
17) \texttt{Dynamo\_256\_Pm10\_Re26} & $ 256^3$ & $7.6$ & $ 10$ & $   26$ & $  260$ & $(  2.6\pm0.3)\!\times\!10^{-1 }$ & $(  4.3\pm1.4)\!\times\!10^{-2 }$ \\
18) \texttt{Dynamo\_128\_Pm10\_Re39} & $ 128^3$ & $8.2$ & $ 10$ & $   39$ & $  390$ & $(  3.4\pm0.3)\!\times\!10^{-1 }$ & $(  4.3\pm1.4)\!\times\!10^{-2 }$ \\
19) \texttt{Dynamo\_256\_Pm10\_Re38} & $ 256^3$ & $8.2$ & $ 10$ & $   38$ & $  380$ & $(  3.2\pm0.3)\!\times\!10^{-1 }$ & $(  5.0\pm1.9)\!\times\!10^{-2 }$ \\
20) \texttt{Dynamo\_512\_Pm10\_Re38} & $ 512^3$ & $8.2$ & $ 10$ & $   38$ & $  380$ & $(  3.1\pm0.6)\!\times\!10^{-1 }$ & n/a \\
21) \texttt{Dynamo\_512\_Pm10\_Re88} & $ 512^3$ & $9.4$ & $ 10$ & $   88$ & $  880$ & $(  4.5\pm0.5)\!\times\!10^{-1 }$ & $(  5.2\pm1.7)\!\times\!10^{-2 }$ \\
22) \texttt{Dynamo\_512\_Pm10\_Re190} & $ 512^3$ & $ 10$ & $ 10$ & $  190$ & $ 1900$ & $(  5.4\pm0.5)\!\times\!10^{-1 }$ & $(  6.0\pm2.0)\!\times\!10^{-2 }$ \\
23) \texttt{Dynamo\_512\_Pm10\_Re390} & $ 512^3$ & $ 10$ & $ 10$ & $  390$ & $ 3900$ & $(  5.9\pm0.6)\!\times\!10^{-1 }$ & $(  6.3\pm2.1)\!\times\!10^{-2 }$ \\
24) \texttt{Dynamo\_256\_Pm10\_Re790} & $ 256^3$ & $ 10$ & $ 10$ & $  790$ & $ 7900$ & $(  6.5\pm0.6)\!\times\!10^{-1 }$ & $(  5.3\pm1.8)\!\times\!10^{-2 }$ \\
25) \texttt{Dynamo\_512\_Pm10\_Re790} & $ 512^3$ & $ 11$ & $ 10$ & $  790$ & $ 7900$ & $(  6.6\pm0.7)\!\times\!10^{-1 }$ & $(  6.4\pm2.1)\!\times\!10^{-2 }$ \\
\hline
\end{tabular*}
\end{table*}

\section{Dynamo theory} \label{sec:theory}

Theories for the turbulent dynamo are based on the \emph{Kazantsev model} \citep{Kazantsev1968,BrandenburgSubramanian2005},
\begin{equation}
  -\kappa_\text{diff}(\ell)\frac{\text{d}^2\psi(\ell)}{\text{d}\ell^2} + U(\ell)\psi(\ell) = -\Gamma \psi(\ell),
\label{eq:Kazantsev}
\end{equation}
which assumes zero helicity, $\delta$-correlation in time, and does not take into account the mixture of solenoidal-to-compressible modes in the turbulent velocity field. These limitations are related to the fact that the Kazantsev equation was historically only applied to incompressible turbulence, while we apply it here to highly compressible, supersonic turbulence.

The similarity of Equation~(\ref{eq:Kazantsev}) with the quantum-mechanical Schr\"odinger equation allows us to solve it both numerically and analytically, which requires an assumption for the scaling of the turbulent velocity correlations. Numerical simulations of turbulence find a power-law scaling within the inertial range ($\ell_\nu < \ell < L$), 
\begin{equation}
\delta v(\ell) \propto \ell^{\vartheta},
\label{eq:v_inertial}
\end{equation}
where $\ell_\nu$ and $L$ are the viscous and integral scale, respectively. The exponent $\vartheta$ varies from $1/3$ for incompressible, non-intermittent Kolmogorov turbulence up to $1/2$ for highly compressible, supersonic Burgers turbulence. Numerical simulations of mildly supersonic turbulence with Mach numbers \mbox{$\mach\sim2$--$7$} find \mbox{$\vartheta\sim0.37$--$0.47$} \citep{BoldyrevNordlundPadoan2002b,KowalLazarian2010,FederrathDuvalKlessenSchmidtMacLow2010}. Highly supersonic turbulence with $\mach>15$ asymptotically approaches the Burgers limit, $\vartheta=0.5$ \citep{Federrath2013}. Observations of interstellar clouds indicate a comparable velocity scaling with \mbox{$\vartheta\sim0.38$--$0.5$} \citep{Larson1981,HeyerBrunt2004,RomanDuvalEtAl2011}. Given this range of exponents, we investigate how the theoretical results depend on $\vartheta$, by studying cases with $\vartheta=0.35$, $0.40$, and $0.45$.

Using the Wentzel-Kramers-Brillouin (WKB) approximation we obtain an analytical solution of the Kazantsev equation, which depends on the velocity scaling exponent $\vartheta$. Results for $\pmag\gg1$ and $\pmag\ll1$ have been reported in \citet{SchoberEtAl2012PRE,SchoberEtAl2012PRE2}. More recently, \citet{BovinoEtAl2013} applied a Numerov scheme to solve Equation~(\ref{eq:Kazantsev}) numerically for \mbox{$\pmag\sim0.1$--$10$}, the regime currently accessible in dynamo simulations. The dependence on the velocity correlation exponent $\vartheta$ forms the main extension of the original, incompressible Kazantsev equation into the compressible regime \citep[note that][have followed a similar approach for mildly-compressible, low-Mach number turbulence]{RogachevskiiKleeorin1997}. However, the generalizations by \citet{SchoberEtAl2012PRE,SchoberEtAl2012PRE2} still do not account for variations in the solenoidal-to-compressible mode mixture that is excited in supersonic turbulence.

\section{Results and Discussion} \label{sec:results}

\begin{figure*}
\centerline{\includegraphics[width=0.99\linewidth]{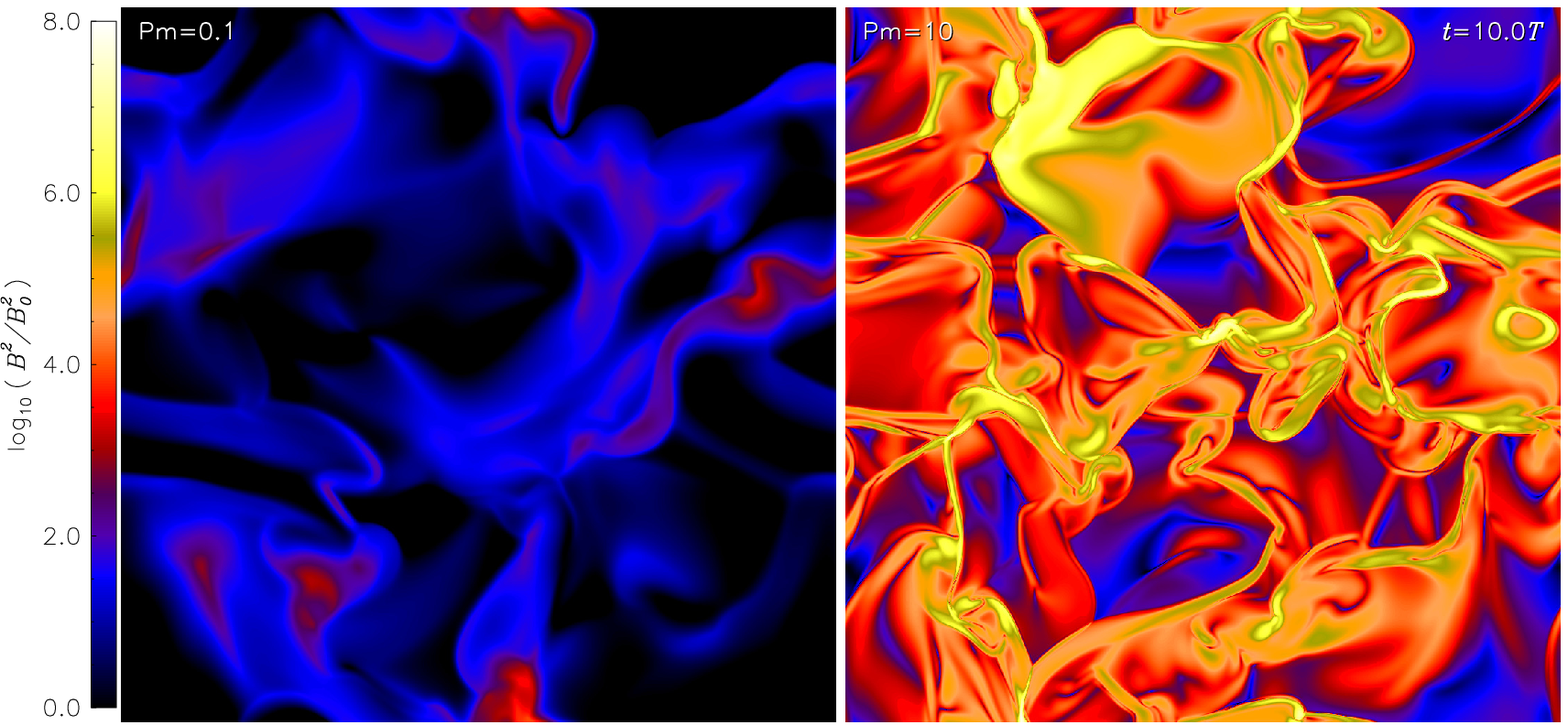}}
\caption{Magnetic energy slices through our simulations with grid resolutions of $1024^3$ points. The magnetic field grows more slowly for magnetic Prandtl numbers of $\pmag=0.1$ (\emph{left-hand panel}) compared to $\pmag=10$ (\emph{right-hand panel}), but we find dynamo action in both cases, for the first time in highly compressible, supersonic plasmas. \emph{An animation of this still shot is available in the online version of the journal.}}
\label{fig:images}
\end{figure*}

To get a visual impression of the differences in the magnetic field structure between low-$\pmag$ and high-$\pmag$ dynamo action, we plot magnetic energy slices in Figure~\ref{fig:images}. By definition, magnetic dissipation is much stronger in low-$\pmag$ compared to high-$\pmag$ turbulence (for $\re=\mathrm{const}$, as in our numerical experiments), but we find that the dynamo operates in both cases. This is the first time that dynamo action is confirmed in low-$\pmag$, highly compressible, supersonic plasma.

\begin{figure*}
\centerline{\includegraphics[width=0.93\linewidth]{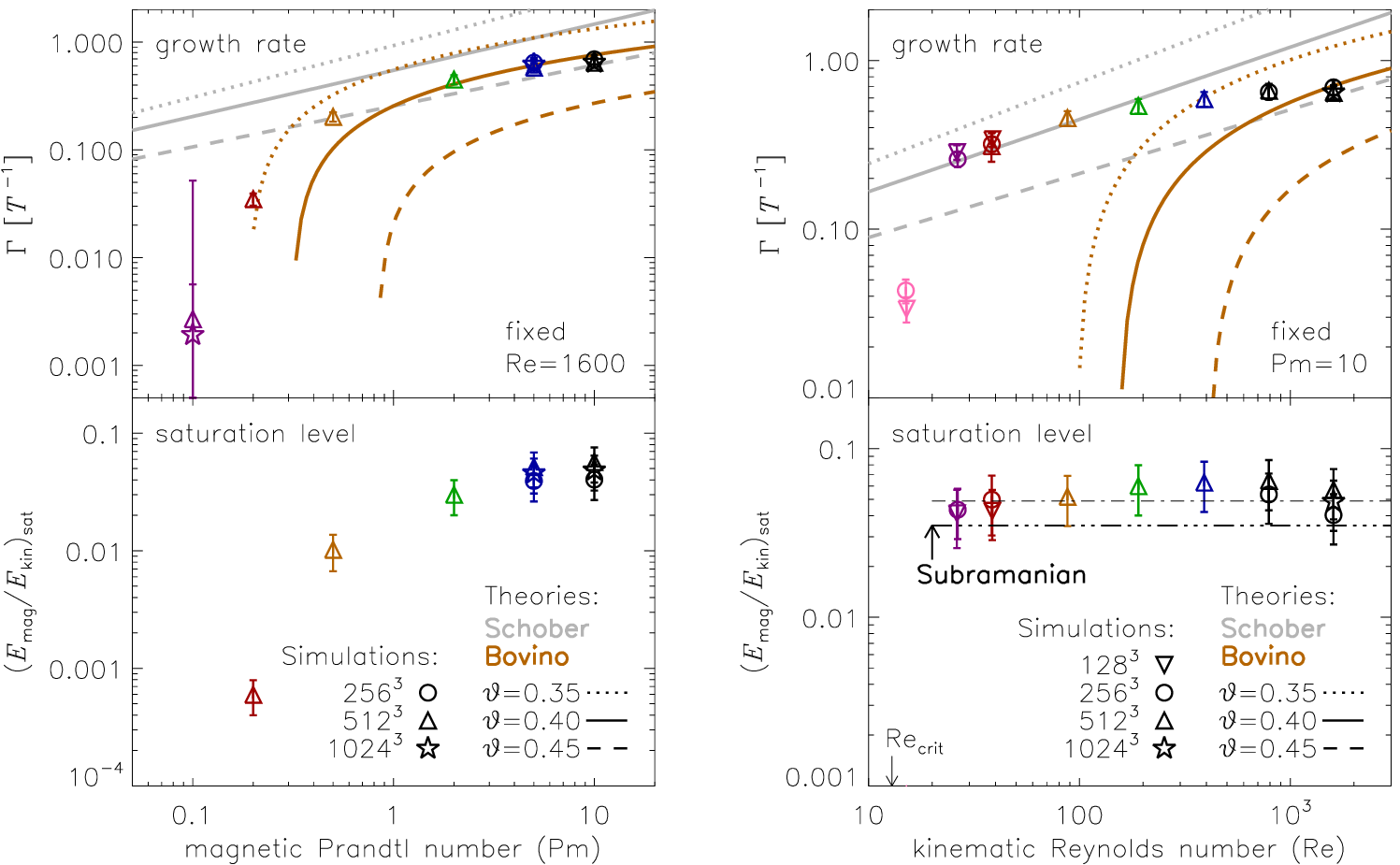}}
\caption{\emph{Left panels:} dynamo growth rate $\Gamma$ (\emph{top panel}) and saturation level $\esat$ (\emph{bottom panel}) as a function of $\pmag$ for fixed $\re=1600$. Resolution studies with $256^3$, $512^3$ and $1024^3$ grid cells demonstrate convergence, tested for the extreme cases $\pmag=0.1$ and $10$. Theoretical predictions for $\Gamma$ by \citet{SchoberEtAl2012PRE2} and \citet{BovinoEtAl2013} are plotted with different line styles for a typical range of the turbulent scaling exponent $\vartheta=0.35$ (dotted), $0.40$ (solid) and $0.45$ (dashed). \emph{Right panels:} same as left panels, but $\Gamma$ and $\esat$ are shown as a function of $\re$ for fixed $\pmag=10$. The dot-dashed line is a fit to the simulations, yielding a constant saturation level of $\esat=0.05\pm0.01$ for $\re>\re_\mathrm{crit}\equiv\rmag_\mathrm{crit}/\pmag=12.9$ and the triple-dot-dashed line shows the result of Subramanian's modified model for the saturation level \citep{Subramanian1999}.}
\label{fig:gratesat}
\end{figure*}

We now determine the dynamo growth rate as a function of $\pmag$ for fixed $\re=1600$ and as a function of $\re$ for fixed $\pmag=10$, in order to compare the analytical and numerical solutions of the Kazantsev equation with the MHD simulations. Depending on $\pmag$ and $\re$ we find exponential magnetic energy growth over more than six orders of magnitude for simulations in which the dynamo is operational. We determine both the exponential growth rate $\Gamma$ and the saturation level $\esat$. The measurements are listed in Table~\ref{tab:sims} and plotted in Figure~\ref{fig:gratesat}.

In the left-hand panel of Figure~\ref{fig:gratesat} we see that $\Gamma$ first increases strongly with $\pmag$ for $\pmag\lesssim1$. For $\pmag\gtrsim1$ it keeps increasing, but more slowly. The theoretical models by \citet{SchoberEtAl2012PRE2} and \citet{BovinoEtAl2013} both predict an increasing growth rate with $\pmag$. The purely analytical solution of the Kazantsev Equation~(\ref{eq:Kazantsev}) by Schober et al., using the WKB approximation, yields power laws for $\rmag>\rmag_\mathrm{crit}$, while the numerical solution of Equation~(\ref{eq:Kazantsev}), using the Numerov method by Bovino et al., yields a sharp cutoff when $\pmag\lesssim1$, closer to the results of the 3D MHD simulations. The agreement of the theoretical prediction with the MHD simulations is excellent for $\pmag\gtrsim1$, while for $\pmag\lesssim1$ they only agree qualitatively. The discrepancy arises because the theoretical models assume zero helicity, $\delta$-correlation of the turbulence in time, and currently do not distinguish different mixtures of solenoidal and compressible modes in the turbulent velocity field. Finite time correlations, however, do not seem to change the Kazantsev result significantly \citep{BhatSubramanian2014} and our simulations have zero helicity. Thus, the missing distinction between solenoidal and compressible modes may be the main cause of the discrepancy, because the dynamo is primarily driven by solenoidal modes and the amount of vorticity strongly depends on the driving and Mach number of the turbulence \citep{MeeBrandenburg2006,FederrathEtAl2011PRL}.

The saturation level as a function of $\pmag$ is shown in the bottom left-hand panel of Figure~\ref{fig:gratesat}. It increases with $\pmag$ similar to the growth rate and is also well converged with increasing numerical resolution. We currently do not have a theoretical model to predict the dynamo saturation level, but it may be possible to develop one based on an effective magnetic diffusivity, which limits the growth of the magnetic field when the back reaction through the Lorentz force prevents turbulence from further stretching, twisting and folding the field \citep{Subramanian1999,BrandenburgSubramanian2005}. However, we currently lack a model that applies to the highly compressible regime of MHD turbulence and that covers the dependence on $\pmag$, although we provide a simple model for the dependence of $\esat$ on $\re$ below.

Finally, the right-hand panels of Figure~\ref{fig:gratesat} show the growth rate and saturation level as a function of $\re$. Similar to the dependence on $\pmag$, we find a non-linear increase in $\Gamma$ with $\re$, which is qualitatively reproduced with the numerical solution by \citet{BovinoEtAl2013}. However, the critical Reynolds number for dynamo action is much lower in the MHD simulations than predicted by the theoretical model, which may have the same reasons as the discrepancy found for the dependence on $\pmag$, i.e., the lack of dependence on the actual turbulent mode mixture in the theoretical models.

In order to determine the critical magnetic Reynolds number for dynamo action, we perform fits with
\begin{equation}
\Gamma = \beta \left[ \ln(\pmag) + \ln(\re) \right] - \gamma,
\label{eq:fit}
\end{equation}
using the fit parameters $\beta$ and $\gamma$, which are related to the critical magnetic Reynolds number $\rmag_\mathrm{crit} = \exp\left(\gamma/\beta\right)$. From the fits with Equation~(\ref{eq:fit}) to all our simulations we find that dynamo action is suppressed for $\rmag < \rmag_\mathrm{crit} = 129^{+43}_{-31}$ in highly compressible, supersonic MHD turbulence. Our result is significantly higher than the critical magnetic Reynolds number measured in simulations of subsonic, incompressible MHD turbulence by \citet{HaugenBrandenburgDobler2004}, who find \mbox{$\rmag_\mathrm{crit}\sim20$--$40$} for $\pmag\gtrsim1$, and higher than in mildly compressible simulations, where $\rmag_\mathrm{crit}\sim50$ for $\pmag=5$ and $\mach\sim2$ \citep{HaugenBrandenburgMee2004}. The reason for the higher $\rmag_\mathrm{crit}$ compared to incompressible turbulence is the more sheet-like than vortex-like structure of supersonic turbulence \citep{Boldyrev2002,SchmidtFederrathKlessen2008} and the reduced fraction of solenoidal modes \citep{MeeBrandenburg2006,FederrathDuvalKlessenSchmidtMacLow2010,FederrathEtAl2011PRL}. The difference with the theoretical models lies primarily in $\rmag_\mathrm{crit}$. \citet{BovinoEtAl2013} predicted a much higher $\rmag_\mathrm{crit}\sim4100$ for $\vartheta=0.45$, while fits to their theoretical model yield \mbox{$\beta=0.11$--$0.19$}, in agreement with the range found in the MHD simulations ($\beta=0.141\pm0.004$). This demonstrates that the discrepancy between the MHD simulations and the Kazantsev model is primarily in the predicted $\rmag_\mathrm{crit}$ value, while the qualitative behavior (determined by the $\beta$ parameter) is covered by the theoretical dynamo models.

The saturation level shown in the bottom right-hand panel of Figure~\ref{fig:gratesat} is consistent with a constant level of $\esat=0.05\pm0.01$ for $\re > \re_\mathrm{crit}\equiv\rmag_\mathrm{crit}/\pmag=12.9$ in highly compressible, supersonic turbulence with Mach numbers $\mach\sim10$, typical for molecular clouds in the Milky Way. Given our measurement of $\rmag_\mathrm{crit}=129$, we can compute Subramanian's theoretical prediction \citep{Subramanian1999} for the saturation level, $\esat=(3/2) (L/V)\tau^{-1}\rmag_\mathrm{crit}^{-1}\sim0.01$, which is significantly smaller than our simulation result, assuming that $\tau=T=L/V$ is the turbulent crossing time on the largest scales of the system. However, Subramanian notes that the timescale $\tau$ is an ``unknown model parameter''. Thus a more appropriate timescale for saturation may be the eddy timescale on the viscous scale, $\ell_\nu = L\re^{-1/(\vartheta+1)}$ for a given turbulent velocity scaling following Equation~(\ref{eq:v_inertial}), because this is where the field saturates first. We find $\tau(\ell_\nu)=\ell_\nu/v(\ell_\nu)=T\re^{(\vartheta-1)/(\vartheta+1)}$ and with $\re=\re_\mathrm{crit}=12.9^{+4.3}_{-3.1}$, we obtain $\esat=0.035\pm0.005$ for a typical range of the velocity scaling exponent $\vartheta=0.4\pm0.1$, from molecular cloud observations and simulations of supersonic turbulence \citep{Larson1981,HeyerBrunt2004,RomanDuvalEtAl2011}. The saturation level of our 3D MHD simulations thus agrees within the uncertainties with our modified version of Subramanian's model. We note that the dependence on $\pmag$ (see the bottom left-hand panel of Figure~\ref{fig:gratesat}) is however not included in the current model and requires further theoretical development.

\begin{figure}
\centerline{\includegraphics[width=0.98\linewidth]{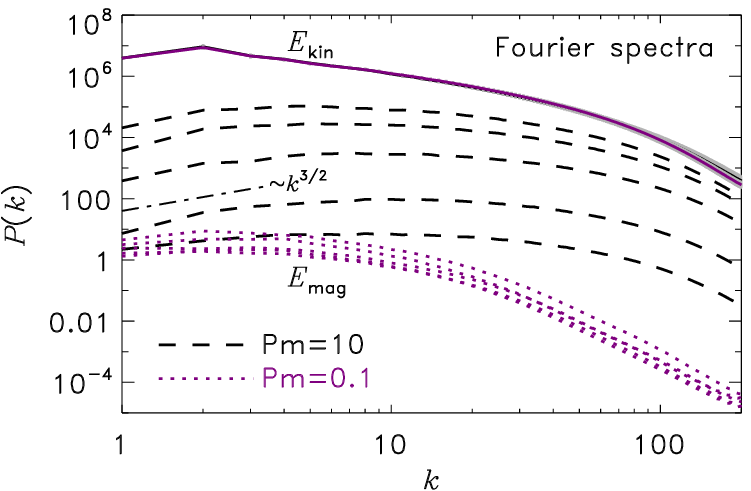}}
\caption{Time evolution of magnetic energy power spectra for simulation models~2 and~11 in Table~\ref{tab:sims}: $\pmag=0.1$ (dotted lines; from bottom to top: $t/T=2$, $5$, $10$, $15$, $18$) and $\pmag=10$ (dashed lines; from bottom to top: $t/T=2$, $5$, $10$, $15$, $24$). The solid lines show the time-averaged kinetic energy spectra with the 1-sigma time variations shown as error bars. Note that for $\pmag=10$, the last magnetic energy spectrum ($t=24\,T$) has just reached saturation on small scales (the $\pmag=0.1$ runs did not reach saturation within the compute time available to us, because the growth rates are so small; cf.~Figure~\ref{fig:gratesat}). The evolution and curvature of the spectra indicate that the magnetic field will continue to grow on large scales during the non-linear dynamo phase.}
\label{fig:spectra}
\end{figure}

To support our conclusions, we show magnetic energy power spectra in Figure~\ref{fig:spectra}. They are qualitatively consistent with the incompressible dynamo studies by \citet{MasonEtAl2011} and \citet{BhatSubramanian2013}. We clearly see that the power spectra for $\pmag=0.1$ dissipate on larger scales (lower $k$) than the $\pmag=10$ spectra, consistent with the theoretical expectation by a factor of \mbox{$(10/0.1)^{1/(1+\theta)}\sim22$--$27$} for our relevant \mbox{$\theta\sim0.4$--$0.5$}. Nevertheless, even for $\pmag=0.1$, we see the dynamo-characteristic increase in magnetic energy over all scales. The magnetic spectra roughly follow the Kazantsev spectrum ($\sim k^{3/2}$) on large scales \citep{Kazantsev1968,BhatSubramanian2014} in the $\pmag=10$ case, but we would expect the same to hold in the $\pmag=0.1$ case, if our simulations had larger scale separation. The final spectrum for $\pmag=10$ has just reached saturation on small scales (approaching the kinetic energy spectrum at high $k$), but continues to grow on larger scales during the non-linear dynamo phase. The $\pmag=0.1$ runs did not have enough time to reach saturation yet (cf.~Figure~\ref{fig:gratesat}), but we expect a qualitatively similar behavior in the non-linear dynamo phase also for models with $\pmag<1$. We emphasize that the kinetic energy spectra shown in Figure~\ref{fig:spectra} and the saturation levels plotted in the bottom panels of Figure~\ref{fig:gratesat} take into account the variations in the density field, i.e., $E_\mathrm{kin}=(1/2)\rho v^2$, because---unlike incompressible turbulence---the density varies by several orders of magnitude in our highly compressible, supersonic turbulence simulations \citep[for a recent analysis of the typical density structures and probability density functions, see][]{Federrath2013}.

\section{Conclusions} \label{sec:conclusions}

We presented the first quantitative comparison of theoretical models of the turbulent dynamo with 3D simulations of supersonic MHD turbulence. We find that the dynamo operates at low and high magnetic Prandtl numbers, but is significantly more efficient for $\pmag>1$ than for $\pmag<1$. We measure a critical magnetic Reynolds number for dynamo action, $\rmag_\mathrm{crit}=129^{+43}_{-31}$ in highly compressible, supersonic turbulence, which is a factor of $\sim3$ times higher than found in studies of subsonic and incompressible turbulence. $\rmag_\mathrm{crit}$ is, however, still several orders of magnitude lower than the magnetic Reynolds number in stars, planets, and in the interstellar medium of galaxies in the present and early Universe, allowing for efficient turbulent dynamo action in all of these environments. This has important consequences for the star formation rate and for the initial mass function of stars, because magnetic fields suppress gas fragmentation and lead to powerful protostellar jets and outflows \citep[see][and references therein]{KrumholzEtAl2014,PadoanEtAl2014,OffnerEtAl2014,FederrathEtAl2014}. We conclude that magnetic fields need to be taken into account during structure formation in the present and early Universe.

\acknowledgements
We thank R.~Banerjee and R.~Klessen for stimulating discussions on the turbulent dynamo and the anonymous referee for their useful comments.
C.F.~acknowledges funding provided by the Australian Research Council's Discovery Projects (grants~DP110102191, DP130102078, DP150104329). D.R.G.S., J.S. and S.B.~thank for funding via the DFG priority program 1573 ``The Physics of the Interstellar Medium'' (grants~SCHL~1964/1-1, BO~4113/1-2).
We gratefully acknowledge the J\"ulich Supercomputing Centre (grant hhd20), the Leibniz Rechenzentrum and the Gauss Centre for Supercomputing (grant pr32lo), the Partnership for Advanced Computing in Europe (PRACE grant pr89mu), and the Australian National Computing Infrastructure (grant ek9).
The software used in this work was in part developed by the DOE-supported Flash Center for Computational Science at the University of Chicago.

\end{document}